\documentclass[12pt]{iopart}
\usepackage[T1]{fontenc}
\usepackage[latin1]{inputenc}
\usepackage[english]{babel}
\usepackage{amssymb}
\usepackage{amsfonts}
\expandafter\let\csname equation*\endcsname\relax
\expandafter\let\csname endequation*\endcsname\relax
\usepackage{amsmath}
\usepackage{iopams}
\usepackage{amsopn}
\usepackage{graphicx}
\usepackage{subfigure}
\usepackage{float,multirow}
\usepackage{color}
\usepackage{ulem}

\begin{document}

\title{Type I superconductivity in Dirac materials}
\author{B. Ya. Shapiro$^{1}$, I. Shapiro$^{1}$, Dingping Li$^{2,3}$ and
Baruch Rosenstein$^{4}$}

\address{$^{1}$Department of Physics, Institute of Superconductivity, Bar-Ilan
University, Ramat-Gan 52900, Israel \\
$^{2}$Collaborative Innovation Center of Quantum Matter, Beijing, China\\
$^{3}$School of Physics, Peking University, Beijing 100871, China\\
$^{4}$Department of Electrophysics, National Chiao Tung University, Hsinchu,
Taiwan, R.O.C. }%
\ead{vortexbar@yahoo.com}

\begin{abstract}
Superconductivity of the second kind was observed in many 3D Weyl and Dirac
semi-metals. However in $PdTe_{2}$, superconductivity is clearly of the
first kind. This is very rare in Dirac semi - metals, but is expected in
clean conventional metallic superconductors with 3D parabolic dispersion
relation. The conduction bands in this material exhibit the linear (Dirac)
dispersion only along two directions, while in the third direction the
dispersion is parabolic. Therefore the "hybrid" Dirac-parabolic material is
intermediate between the two extremes. A microscopic pairing theory is
derived for arbitrary tilt parameter of the 2D cone and used to determine
anisotropic coherence lengths, the penetration depths and applied to recent
extensive experiments. Magnetic properties of these superconductors are then
studied on the basis of microscopically derived Ginzburg - Landau effective
theory for the order parameter.
\end{abstract}

\maketitle

\section{Introduction}

Dispersion relation near Fermi surface in recently synthesized two and three
dimensional Dirac (Weyl) semi-metals\cite{Weng} ,\cite{MoTeearly},\cite{ZrTe}
is linear, qualitatively distinct from conventional metals, semi - metals or
semiconductors in which it is parabolic. In type I Dirac semi-metals (DSM),
the band inversion results in Dirac points in low-energy excitations being
anisotropic massless "relativistic" fermions. More recently type-II DSM with
Dirac cone strongly tilted, so that they can be characterized by a nearly
flat band at Fermi surface were discovered\cite{Soluyanov}. The type-II DSM
also exhibit exotic properties different from the type-I ones. Many Dirac
materials are known to be superconducting. A detailed study of
superconductivity in DSM under hydrostatic pressure revealed a curious
dependence of critical temperature of the superconducting transition on
pressure. The critical temperature $T_{c\text{ }}$ in some of these systems
like $HfTe_{5}$ shows\cite{HfTe} a maximum as a function of pressure.
Superconductivity happens to be of the second kind with penetration depth $%
\lambda $ much larger than the coherence length $\xi $. However in recently
studied material \cite{PdTe2} $PdTe_{2}$ it was demonstrated that
superconductivity is of the first kind.

Although various pairing mechanisms in DSM turned superconductors have been
considered \cite{DasSarma,FuBerg,frontiers}, experiments indicate the
conventional phonon mediated one. If the Fermi level is not situated too
close to the Dirac point, the BCS type pairing occurs, otherwise a more
delicate formalism should be employed\cite{Shapiro14}. A theory predicted
possibility of superconductivity in the type II Weyl semimetals was
developed recently in the framework of Eliashberg model \cite%
{Zyuzin,Rosenstein17}. In particular the case strongly layered 2D Dirac
materials in clean limit like $MoTe_{2}$\cite{MoTe2melting} was considered
in \cite{Li18}. The critical fields, coherence lengths magnetic penetration
depths and the Ginzburg number characterizing the strength of fluctuations
were found. It turned out that in most cases the superconductivity was of
the second kind. Moreover the thermal fluctuations were shown to be strong
enough to qualitatively affect the Abrikosov vortex phase diagram. The
vortex lattice "melts" into the vortex liquid. This is reminiscent of a well
known (possibly non - Dirac semi-metal) layered dichalcogenides
superconductor $NbSe_{2}$ that is perhaps the only low $T_{c}$ material with
fluctuations strong enough to exhibit vortex lattice melting\cite{NbSe2}.

The layered superconductor is similar to the present case in that the Dirac
spectrum is two dimensional and parabolic in the third direction. However it
is qualitatively different in that the anisotropy in the third direction is
extremely strong. In $PdTe_{2}$ the dispersion relation is parabolic, but
anisotropy is mild. Thus the "hybrid" Dirac-parabolic materials can be
viewed as an intermediate between the two extremes, 3D DSM and conventional
metals in the "clean limit"\cite{Ketterson}.

Superconductivity in $PdTe_{2}$ with a transition temperature $T_{c}$ of $%
1.5K$ was discovered in 1961 by Guggenheim et al.\cite{Guddenheim}. The
material was revisited recently when the type II Dirac dispersion relation
was observed by ARPES \cite{Noh}. In this material, the pair of type-II
Dirac points disappears at $6.1GPa$, while a new pair of type-I Dirac points
emerges at $4.7GPa$. It was recently predicted by theories and confirmed in
experiments, making $PdTe_{2}$ \cite{PdTe2} the first material that
processes both superconductivity and type-II Dirac fermions under proper
pressure ($4.7-6.1GPa$). An early determination of $T_{c}$ was confirmed by
others with $T_{c}$ values ranging from $1.7$ to $2.0K$ \cite{Fei}. While
other 3D Dirac semi-metal $ZrTe_{5}$ $Na_{3}Bi$ \cite{NaBi} and $%
Cd_{3}As_{2} $ \cite{CdAs} demonstrates magnetic properties typical for
second kind superconductivity, the magnetic and transport measurements on
the single crystals unambiguously show that $PdTe_{2}$ is a first kind
superconductor \cite{Leng}. It makes $PdTe_{2}$ system to be the first
Dirac/Weyl semimetal where superconductivity is of first kind.

In the present paper we extend the study of superconductivity in the
"hybrid" Dirac-parabolic clean semimetals. The phenomenological
Ginzburg-Landau theory for superconducting DSM of the arbitrary type is
microscopically derived and used to establish magnetic phase diagram. In
particular the Abrikosov parameter $\kappa ^{A}$ used to distinguish between
the superconductivity of the first from the second kind is determined. We
applied our theory to explanation of the recent studied material, $PdTe_{2}$
as a representative example of hybrid layered DSM. A major reason is that
magnetic properties of this superconductor were investigated in a wide range
of temperatures and magnetic fields with the magnetic field was directed
parallel to the layers. An additional advantage of this choice is that the
material $PdTe_{2}$ in many aspects behaves as a 3D anisotropic material.

In the present paper a microscopic pairing theory is constructed and used to
determine anisotropic coherence lengths, the penetration depths,
thermodynamic critical field. The results are applied to recent extensive
experiments on $PdTe_{2}$. Magnetic properties of these superconductors are
studied on the basis of microscopically derived Ginzburg - Landau effective
theory for the order parameter.

The paper is organized as follows. In section II a sufficiently general
phonon mediated BCS model of the "hybrid" type I and II DSM is formulated.
Gor'kov equations are written with details relegated to appendices. The
section III contains derivation of the coefficients of the Ginzburg - Landau
equations from the Gor'kov equations in the inhomogeneous case. Magnetic
properties are derived from the GL model in section IV.

\section{Model and Gorkov's equations}

The band structure near the Fermi level of a DSM is well captured by the
non-interacting massless Weyl Hamiltonian with the Fermi velocity $v$
(assumed to be isotropic in the $x-y$ plane) and conventional parabolic term
on $z-$direction Fig.1. \cite{Rosenstein17},\cite{Li18}:

\begin{eqnarray}
K &=&\int_{\mathbf{r}}\psi _{\alpha }^{s+}\left( \mathbf{r}\right) \widehat{K%
}_{\alpha \beta }\psi _{\beta }^{s}\left( \mathbf{r}\right) \text{\ \ \ \ }
\label{eq1} \\
\text{\ \ }\widehat{K}_{\gamma \delta } &=&-i\hbar v\nabla ^{i}\sigma
_{\gamma \delta }^{i}+\left( -i\hbar w_{i}\nabla ^{i}-\mu +\frac{p_{z}^{2}}{%
2m_{z}}\right) \delta _{\gamma \delta }\text{.}  \notag
\end{eqnarray}%
Here $\mu $ is the chemical potential, $p_{z}=-i\hbar \nabla _{z}$ , $\sigma
$ are Pauli matrices in the sublattice space in the WSM layers, with just
two sublattices denoted by $\alpha =1,2,$ and $s$ is spin projection. The
velocity vector $\mathbf{w}$ defines the tilt of the (otherwise isotropic)
cone. The graphene - like dispersion relation for $\mathbf{w}=0$ represents
the type I Weyl semi-metal, while for the velocity $\left\vert \mathbf{w}%
\right\vert =w$ exceeding $v$, the material becomes a type II Weyl semi -
metal.

\begin{figure}[t]
\centering
\includegraphics[width=12cm]{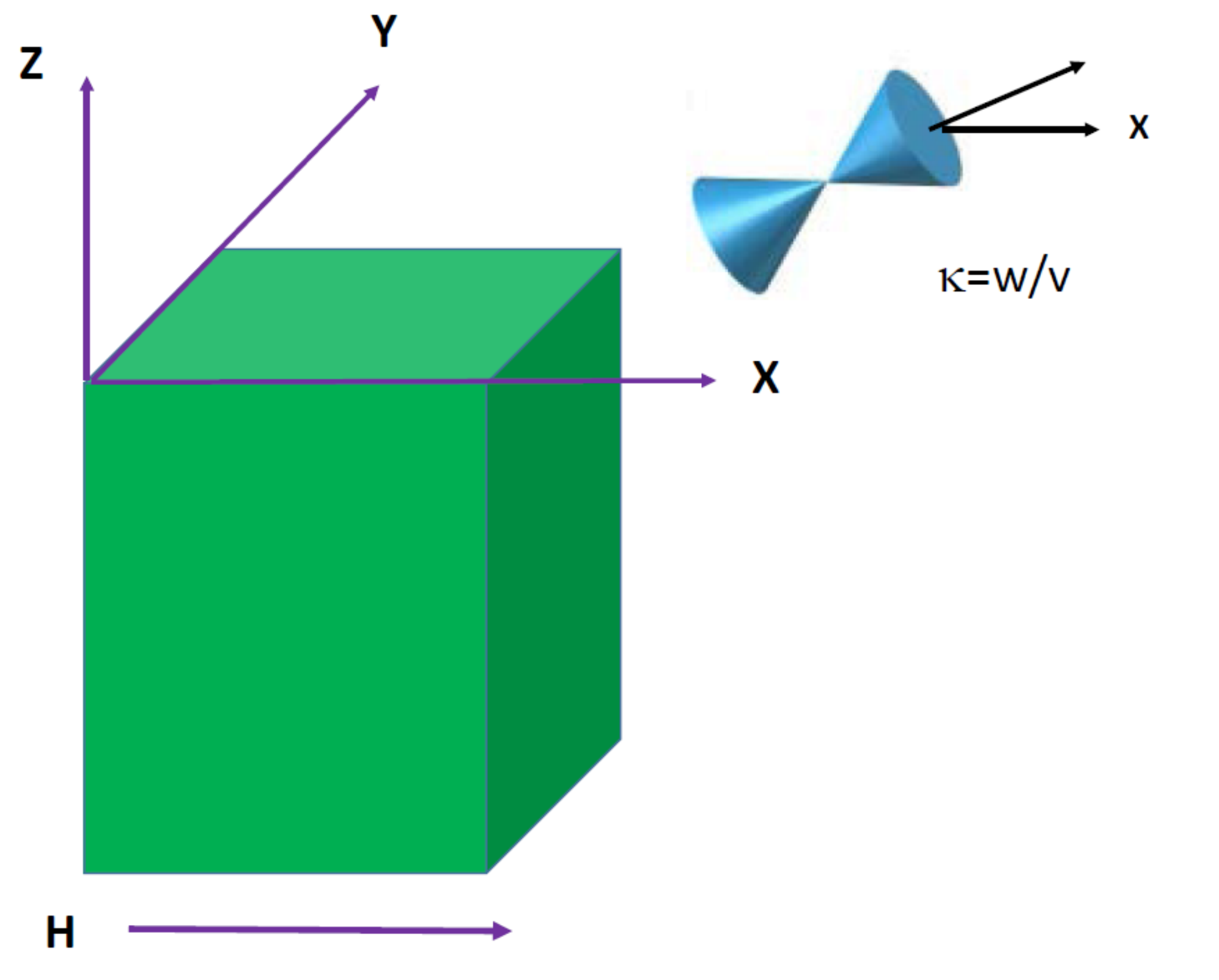} \vspace{-0.5cm}
\caption{ Geometry of DSM sample. Dirac dispersion relation in the $x-y$
plane while parabolic dispersion law is in $z-$direction. Two dimensional
Dirac cone is tilted in x-axis direction with cone tilted parameter $\protect%
\kappa =w/v$ (here  Fermi velocity $v$ is assumed to be isotropic in the $x-y
$ plane).  }
\end{figure}

Generally there are a number of pairs of points (Weyl cones) constituting
the Fermi "surface" of such a material at chemical potential $\mu =0$. We
restrict ourself to the case of just one left handed and one right handed
Dirac points, typically but not always separated in the Brillouin zone.
Generalization to include the opposite chirality and several "cones" is
straightforward. We assume that different valleys are paired independently
and drop the valley indices (multiplying the density of states by number of
valleys).

The effective electron-electron attraction due to the electron - phonon
interaction opposed by Coulomb repulsion (pseudopotential) mechanism creates
pairing below $T_{c}$. For simplicity we assume the singlet $s$-channel
interaction with essentially local interaction,
\begin{equation}
V=\frac{g^{2}}{2}\ \int d\mathbf{r}\text{ }\psi _{\alpha }^{+\uparrow
}\left( \mathbf{r}\right) \psi _{\beta }^{\downarrow +}\left( \mathbf{r}%
\right) \psi _{\beta }^{\uparrow }\left( \mathbf{r}\right) \psi _{\alpha
}^{\downarrow }\left( \mathbf{r}\right) \text{,}  \label{int}
\end{equation}%
where the coupling $g^{2}$ is zero between the layers. As usual the retarded
interaction has a cutoff frequency $\Omega $, so that it is active in an
energy shell of width $2\hbar \Omega $ around the Fermi level \cite%
{Abrikosov}. For the phonon mechanism it is the Debye frequency. We first
remind\cite{Rosenstein17}, the Gorkov equations and then derive from them
the phenomenological GL equations that allow to obtain the basic magnetic
response of the superconductors.

Finite temperature properties of the condensate are described at temperature
$T$ by the normal and the anomalous Matsubara Greens functions\cite%
{Abrikosov} (GF),%
\begin{eqnarray}
G_{\alpha \beta }^{ts}\left( \mathbf{r}\tau ,\mathbf{r}^{\prime }\tau
^{\prime }\right) &=&-\left\langle T_{\tau }\psi _{\alpha }^{t}\left(
\mathbf{r}\tau \right) \psi _{\beta }^{s+}\left( \mathbf{r}^{\prime }\tau
^{\prime }\right) \right\rangle =\delta ^{ts}g_{\alpha \beta }\left( \mathbf{%
r-r}^{\prime },\tau -\tau ^{\prime }\right) ;  \label{green} \\
F_{\alpha \beta }^{ts}\left( \mathbf{r}\tau ,\mathbf{r}^{\prime }\tau
^{\prime }\right) &=&\left\langle T_{\tau }\psi _{\alpha }^{t}\left( \mathbf{%
r}\tau \right) \psi _{\beta }^{s}\left( \mathbf{r}^{\prime }\tau ^{\prime
}\right) \right\rangle =-\varepsilon ^{ts}f_{\alpha \beta }\left( \mathbf{r-r%
}^{\prime },\tau -\tau ^{\prime }\right) ;  \notag \\
F_{\alpha \beta }^{+ts}\left( \mathbf{r}\tau ,\mathbf{r}^{\prime }\tau
^{\prime }\right) &=&\left\langle T_{\tau }\psi _{\alpha }^{t+}\left(
\mathbf{r}\tau \right) \psi _{\beta }^{s+}\left( \mathbf{r}^{\prime }\tau
^{\prime }\right) \right\rangle =\varepsilon ^{ts}f_{\alpha \beta
}^{+}\left( \mathbf{r-r}^{\prime },\tau -\tau ^{\prime }\right) .  \notag
\end{eqnarray}%
where $t,s$ are the spin indexes. The set of Gor'kov equations in the time
translation invariant, yet inhomogeneous case is\cite{Rosenstein17,Li18},

\begin{eqnarray}
L_{\gamma \beta }^{1}g_{\beta \kappa }\left( \mathbf{r,r}^{\prime }\ \omega
\right) &=&\delta ^{\gamma \kappa }\delta \left( \mathbf{r-r}^{\prime
}\right) -\Delta _{\alpha \gamma }\left( \mathbf{r,}\tau =0\right) f_{\alpha
\kappa }^{+}\left( \mathbf{r,r}^{\prime },\omega \right) ;  \label{FGE} \\
L_{\gamma \beta }^{2}f_{\beta \kappa }^{+}\left( \mathbf{r,r}^{\prime
},\omega \right) &=&\Delta _{\beta \gamma }^{\ast }\left( \mathbf{r,}\tau
=0\right) g_{\beta \kappa }\left( \mathbf{r,r}^{\prime },\omega \right)
\text{.}  \notag
\end{eqnarray}%
Here the two Weyl operators are, (tilt vector $\mathbf{w}$ is assumed to be
directed along $x$- axes):%
\begin{eqnarray}
L_{\gamma \beta }^{1} &=&\left[ \left( i\omega +\mu ^{\prime }+iw\nabla
_{x}\right) \delta _{\gamma \beta }-iv\sigma _{\gamma \beta }^{i}\ \nabla
_{r}^{i}\right] ;  \label{Eq.12} \\
L_{\gamma \beta }^{2} &=&\left[ \left( -i\omega +\mu ^{\prime }+iw\nabla
_{x}\right) \delta _{\gamma \beta }-iv\sigma _{\gamma \beta }^{it}\nabla
_{r}^{i}\right] \text{,}  \notag
\end{eqnarray}%
The effective 2D chemical potential was denoted by $\mu ^{\prime }\equiv \mu
-\frac{p_{z}^{2}}{2m_{z}}$.

The gap function defined as\bigskip
\begin{equation}
\Delta _{\beta \kappa }^{\ast }\left( \mathbf{r}\right)
=g^{2}T\sum\limits_{\omega }f_{\beta \kappa }^{+}\left( \mathbf{r,}\omega
\right) .  \label{Eq. 20}
\end{equation}%
The gap function in the s-wave channel is $\Delta _{\alpha \gamma }\left(
\mathbf{r}\right) =\sigma _{\alpha \gamma }^{x}\Delta \left( \mathbf{r}%
\right) .$ This is the starting point for derivation of the GL free energy
functional of $\Delta \left( \mathbf{r}\right) $.

\section{Derivation of the GL equations}

In this section the Ginzburg - Landau equations in a homogeneous material
(including the gradient terms) is derived. Magnetic field and fluctuations
effects will be discussed in the next two section by generalizing the basic
formalism. To derive the GL equations including the derivative term one
needs the integral form of the Gor'kov equations (see Appendix A), Eq.(\ref%
{FGE}):
\begin{eqnarray}
g_{\epsilon \kappa }\left( \mathbf{r,r}^{\prime },\omega \right) &=&\mathrm{g%
}_{\epsilon \kappa }^{1}\left( \mathbf{r}-\mathbf{r}^{\prime },\omega
\right) -\int_{\mathbf{r}^{\prime \prime }}\mathrm{g}_{\epsilon \theta
}^{1}\left( \mathbf{r}-\mathbf{r}^{\prime \prime },\omega \right) \Delta
_{\theta \phi }^{\ast }\left( \mathbf{r}^{\prime \prime }\right) f_{\phi
\kappa }^{+}\left( \mathbf{r}^{\prime \prime }\mathbf{,r}^{\prime },\omega
\right) ;  \label{IE1} \\
f_{\beta \kappa }^{+}\left( \mathbf{r,r}^{\prime },\omega \right) &=&\int_{%
\mathbf{r}^{\prime \prime \prime }}\mathrm{g}_{\beta \alpha }^{2}\left(
\mathbf{r-r}^{\prime \prime \prime },-\omega \right) \Delta _{\alpha
\epsilon }^{\ast }\left( \mathbf{r}^{\prime \prime \prime }\right) \times
\notag \\
&&\left\{ \mathrm{g}_{\epsilon \kappa }^{1}\left( \mathbf{r}^{\prime \prime
\prime }-\mathbf{r}^{\prime },\omega \right) -\int_{\mathbf{r}^{\prime
\prime }}\mathrm{g}_{\epsilon \theta }^{1}\left( \mathbf{r}^{\prime \prime }-%
\mathbf{r}^{\prime \prime \prime },\omega \right) \Delta _{\theta \phi
}^{\ast }\left( \mathbf{r}^{\prime \prime }\right) f_{\phi \kappa
}^{+}\left( \mathbf{r}^{\prime \prime }\mathbf{,r}^{\prime },\omega \right)
\right\} \text{.}  \notag
\end{eqnarray}%
Here $\mathrm{g}_{\beta \kappa }^{1}\left( \mathbf{r,r}^{\prime }\right) $
and $\mathrm{g}_{\beta \kappa }^{2}\left( \mathbf{r,r}^{\prime }\right) $
are GF of operators $L_{\gamma \beta }^{1}$ and $L_{\gamma \beta }^{2}$:

\begin{equation}
L_{\gamma \beta }^{1}\mathrm{g}_{\beta \kappa }^{1}\left( \mathbf{r,r}%
^{\prime }\right) =\delta ^{\gamma \kappa }\delta \left( \mathbf{r-r}%
^{\prime }\right) ;L_{\gamma \beta }^{2}\mathrm{g}_{\beta \kappa }^{2}\left(
\mathbf{r,r}^{\prime }\right) =\delta ^{\gamma \kappa }\delta \left( \mathbf{%
r-r}^{\prime }\right) .  \label{Eq.14}
\end{equation}%
This will be enough do derive the GL expansion to the third order in the gap
function $\Delta \left( \mathbf{r}\right) $ that will be used as an order
parameter\cite{Abrikosov}.

Using the first and the second iteration of equations Eq.(\ref{IE1}) and
specializing on the case $\mathbf{r}=\mathbf{r}^{\prime }$, and specify the
Fourier the Fourier transformation for the GF,

\begin{equation}
\mathrm{g}_{\alpha \beta }^{2,1}\left( \mathbf{r}\right) =\sum\nolimits_{%
\mathbf{p}}g_{\alpha \beta }^{2,1}\left( \mathbf{p}\right) e^{i\mathbf{%
p\cdot r}},\Delta \left( \mathbf{r}\right) =\sum\nolimits_{\mathbf{q}}\Delta
\left( \mathbf{q}\right) e^{i\mathbf{q\cdot r}}
\end{equation}%
one rewrites the Gorkov's equation Eq.(\ref{FGE}) as (see details in Ref.
\cite{Li18}):,
\begin{equation}
\Delta \left( \mathbf{r}\right) =\frac{g^{2}T}{2}\sum\limits_{\omega ,%
\mathbf{p}}\ \left\{ a\left( \mathbf{p}\right) \Delta \left( \mathbf{r}%
\right) \mathbf{+}C_{ki}\left( \mathbf{p}\right) \frac{\partial ^{2}\Delta
\left( \mathbf{r}\right) }{\partial \mathbf{r}_{i}\partial \mathbf{r}_{k}}%
-b\left( \mathbf{p}\right) \Delta ^{3}\left( \mathbf{r}\right) \right\}
\text{.}  \label{GL17}
\end{equation}%
The function appearing in an expression for the coefficient $a$ is:

\begin{equation}
a\left( \mathbf{p}\right) =g_{21}^{2}\left( \mathbf{p}\right)
g_{21}^{1}\left( \mathbf{p}\right) +g_{11}^{2}\left( \mathbf{p}\right)
g_{22}^{1}\left( \mathbf{p}\right) +g_{12}^{2}\left( \mathbf{p}\right)
g_{12}^{1}\left( \mathbf{p}\right) +g_{22}^{2}\left( \mathbf{p}\right)
g_{11}^{1}\left( \mathbf{p}\right) \text{,}  \label{Eq. 38}
\end{equation}%
while the gradient term coefficients take a form:%
\begin{equation}
C_{ki}\left( \mathbf{p}\right) =\frac{1}{2}\left\{
\begin{array}{c}
\frac{\partial g_{21}^{2}\left( \mathbf{p}\right) }{\partial p_{k}}\frac{%
\partial g_{21}^{1}\left( \mathbf{p}\right) }{\partial p_{i}}+\frac{\partial
g_{11}^{2}\left( \mathbf{p}\right) }{\partial p_{k}}\frac{\partial
g_{22}^{1}\left( \mathbf{p}\right) }{\partial p_{i}}+ \\
\frac{\partial g_{12}^{2}\left( \mathbf{p}\right) }{\partial p_{k}}\frac{%
\partial g_{12}^{1}\left( \mathbf{p}\right) }{\partial p_{i}}+\frac{\partial
g_{22}^{2}\left( \mathbf{p}\right) }{\partial p_{k}}\frac{\partial
g_{11}^{1}\left( \mathbf{p}\right) }{\partial p_{i}}%
\end{array}%
\right\} \text{.}  \label{Eq. 39}
\end{equation}%
The cubic term's coefficient is given by%
\begin{equation}
b\left( \mathbf{p}\right) =\left\{
\begin{array}{c}
g_{21}^{2}\left( \mathbf{p}\right) g_{22}^{1}\left( -\mathbf{p}\right)
g_{11}^{2}\left( -\mathbf{p}\right) g_{21}^{1}\left( \mathbf{p}\right)
+g_{21}^{2}\left( \mathbf{p}\right) g_{21}^{1}\left( -\mathbf{p}\right)
g_{21}^{2}\left( -\mathbf{p}\right) g_{21}^{1}\left( \mathbf{p}\right) + \\
g_{22}^{2}\left( \mathbf{p}\right) g_{11}^{1}\left( -\mathbf{p}\right)
g_{22}^{2}\left( -\mathbf{p}\right) g_{11}^{1}\left( \mathbf{p}\right)
+g_{22}^{2}\left( \mathbf{p}\right) g_{12}^{1}\left( -\mathbf{p}\right)
g_{12}^{2}\left( -\mathbf{p}\right) g_{11}^{1}\left( \mathbf{p}\right) + \\
g_{11}^{2}\left( \mathbf{p}\right) g_{21}^{1}\left( -\mathbf{p}\right)
g_{21}^{2}\left( -\mathbf{p}\right) g_{22}^{1}\left( \mathbf{p}\right)
+g_{11}^{2}\left( \mathbf{p}\right) g_{22}^{1}\left( -\mathbf{p}\right)
g_{11}^{2}\left( -\mathbf{p}\right) g_{22}^{1}\left( \mathbf{p}\right) \ +
\\
g_{12}^{2}\left( \mathbf{p}\right) g_{11}^{1}\left( -\mathbf{p}\right)
g_{22}^{2}\left( -\mathbf{p}\right) g_{12}^{1}\left( \mathbf{p}\right)
+g_{12}^{2}\left( \mathbf{p}\right) g_{12}^{1}\left( -\mathbf{p}\right)
g_{12}^{2}\left( -\mathbf{p}\right) g_{12}^{1}\left( \mathbf{p}\right)%
\end{array}%
\right\} \text{.}  \label{Eq.40}
\end{equation}%
\ .

Normal Green function are obtained\cite{Li18} from equations Eq.(\ref{Eq.14}%
):

\begin{eqnarray}
g_{22}^{1}\left( \mathbf{p}\right) &=&z^{\ast -1}\left( i\omega +\mu
^{\prime }-\mathbf{wp}\right) ;\text{ \ }g_{12}^{1}\left( \mathbf{p}\right)
=-z^{\ast -1}vpe^{-i\varphi };  \label{GF} \\
g_{11}^{1}\left( \mathbf{p}\right) &=&z^{\ast -1}\left( i\omega +\mu
^{\prime }-\ \mathbf{wp}\right) ;\text{ }g_{21}^{1}\left( \mathbf{p}\right)
=-z^{\ast -1}vpe^{i\varphi };  \notag \\
g_{11}^{2}\left( \mathbf{p}\right) \ &=&z^{-1}\ \left( -i\omega +\mu
^{\prime }-\mathbf{wp}\right) ;g_{12}^{2}\left( \mathbf{p}\right)
=-z^{-1}vpe^{i\varphi };  \notag \\
g_{22}^{2}\left( \mathbf{p}\right) &=&z^{-1}\left( -i\omega +\mu ^{\prime
}-\ \mathbf{wp}\right) ;\text{ \ }g_{21}^{2}\left( \mathbf{p}\right)
=-z^{-1}vpe^{-i\varphi }\ .  \notag
\end{eqnarray}%
See Appendix B for details. \bigskip\ Here $z\equiv \left( -i\omega +\mu
^{\prime }-\mathbf{wp}\right) ^{2}-\left( vp\right) ^{2},\mathbf{p}$ is the
2D momentum and $\varphi $ is the azimuthal angle in the $p_{x},p_{y}$ plane.

\section{Critical temperature and the linear term in GL expansion.}

\subsection{Critical temperature}

The linear terms in the GL expansion read:

\begin{equation}
a\left( T\right) =T\sum\nolimits_{\omega ,\mathbf{p}}a\left( \mathbf{p}%
\right) -\frac{1}{g^{2}},  \label{Eq.42}
\end{equation}%
while the critical temperature $T_{c}$ is defined by the condition $a\left(
T_{c}\right) =0$. Substituting GF of Eqs.(\ref{GF}) into Eq.(\ref{Eq. 38}),
one obtains in dimensionless variables, $\overline{\omega }=\pi \left(
2n+1\right) ,\varepsilon =vp/T,\varepsilon _{z}=p_{z}^{2}/2m_{z}T,\overline{%
\mu }=\mu /T$,
\begin{equation}
a\left( T\right) =\frac{1}{\lambda }-\frac{3}{16\pi f}\frac{1}{\overline{\mu
}^{3/2}}\sum\limits_{n}\int_{shell}\frac{\varepsilon d\varepsilon
_{z}d\varphi d\varepsilon }{\sqrt{\varepsilon _{z}}}\frac{\ \overline{\omega
}^{2}+\varepsilon ^{2}+\Phi ^{2}}{\ \left( \overline{\omega }^{2}+\left(
\Phi -\varepsilon \right) ^{2}\right) \left( \ \overline{\omega }^{2}+\left(
\ \Phi +\varepsilon \right) ^{2}\right) }\text{.}  \label{Tc(k)}
\end{equation}%
Here $\Phi =\left( \overline{\mu }-\varepsilon _{z}-\kappa \varepsilon \cos
\varphi \right) ,\lambda =g^{2}D\left( \mu \right) =\lambda _{0}f$, and the
density of states of the normal state electrons (per spin and valley) is
\begin{equation}
D\left( \mu \right) =\frac{\sqrt{2m_{z}}\mu ^{3/2}}{12\pi ^{2}\hbar ^{3}v^{2}%
}f\left( \kappa \right) \text{.}  \label{DOS}
\end{equation}%
Dimensionless constant $\lambda _{0}$ is the electron-electron strength for
zero tilt parameter $\kappa $.

The function
\begin{equation}
f\left( \kappa \right) =\frac{1}{2\pi }\int\limits_{\varphi =0}^{\pi }\frac{%
sign\left( \kappa \cos \varphi +1\right) }{\left( \kappa \cos \varphi
+1\right) ^{2}}.  \label{f}
\end{equation}%
is different for Type I and Type II DSM, but has the same form as in 2D DSM%
\cite{Rosenstein17},\cite{Li18}. For the type I WSM, $\kappa <1$, in which
the Fermi surface is a closed ellipsoid, it is given by:%
\begin{equation}
f=\frac{1}{\left( 1-\kappa ^{2}\right) ^{3/2}}\text{.}  \label{fI}
\end{equation}%
In the type II phase, $\kappa >1$, the Fermi surface becomes open, extending
over the Brillouin zone, and the corresponding expression is:%
\begin{equation}
f=\frac{\kappa ^{2}}{\ \pi \left( \kappa ^{2}-1\right) ^{3/2}}\left\{ 2\sqrt{%
1+\kappa }-1+\log \left[ \frac{2\left( \kappa ^{2}-1\right) }{\kappa \left(
1+\sqrt{1+\kappa }\right) ^{2}\delta }\right] \right\} \text{.}  \label{fII}
\end{equation}%
The integration in Eq.(\ref{Tc(k)}) is performed in the BCS shell around the
chemical potential:\bigskip
\begin{equation}
\varepsilon +\varepsilon _{z}+\overline{\Omega }>\overline{\mu }>\varepsilon
+\varepsilon _{z}-\overline{\Omega }\text{.}  \label{shell}
\end{equation}

\bigskip After Matsubara frequencies summation one obtains,

\begin{equation}
a\left( T\right) =\frac{1}{\lambda }-\ \frac{3}{8}\frac{1}{\overline{\mu }%
^{3/2}}\int_{shell}\frac{d\varepsilon _{z}}{\sqrt{\varepsilon _{z}}}%
EdE\left( \frac{\tanh \frac{\varepsilon _{z}+E-\overline{\mu }}{2}}{%
\varepsilon _{z}+E-\overline{\mu }}+\frac{\tanh \frac{\left( \varepsilon
_{z}-E-\overline{\mu }\right) }{2}}{\varepsilon _{z}-E-\overline{\mu }}%
\right) ,  \label{lamda}
\end{equation}%
where $E=\kappa \varepsilon \cos \varphi +\varepsilon $. Performing
integration in Eq.(\ref{lamda}), one obtains for $T_{c}$ and $a\left(
T\right) $ usual BCS expressions,$\ \ $%
\begin{equation}
T_{c}=1.14\Omega \exp \left( -\frac{1}{f\left( \kappa \right) \lambda _{0}}%
\right)  \label{Tc3D}
\end{equation}%
\begin{equation*}
a\left( T\right) \simeq \frac{T_{c}-T}{T_{c}}
\end{equation*}%
The critical temperature as a function of the cone tilt parameter $\kappa $
is presented in Fig.2.

\bigskip

Fig.2 Critical superconducting temperature as a function on cone tilt
parameter $\kappa $ plotted for electron-electron strength $\lambda
_{0}=0.2, $ $m_{z}=m_{e},\mu =25\Omega ,\Omega =100K.$ Green dash line marks
the critical temperature of the $PdTe_{2}$ sample from Ref. \cite{Leng}.

\begin{figure}[t]
\centering
\includegraphics[width=14cm]{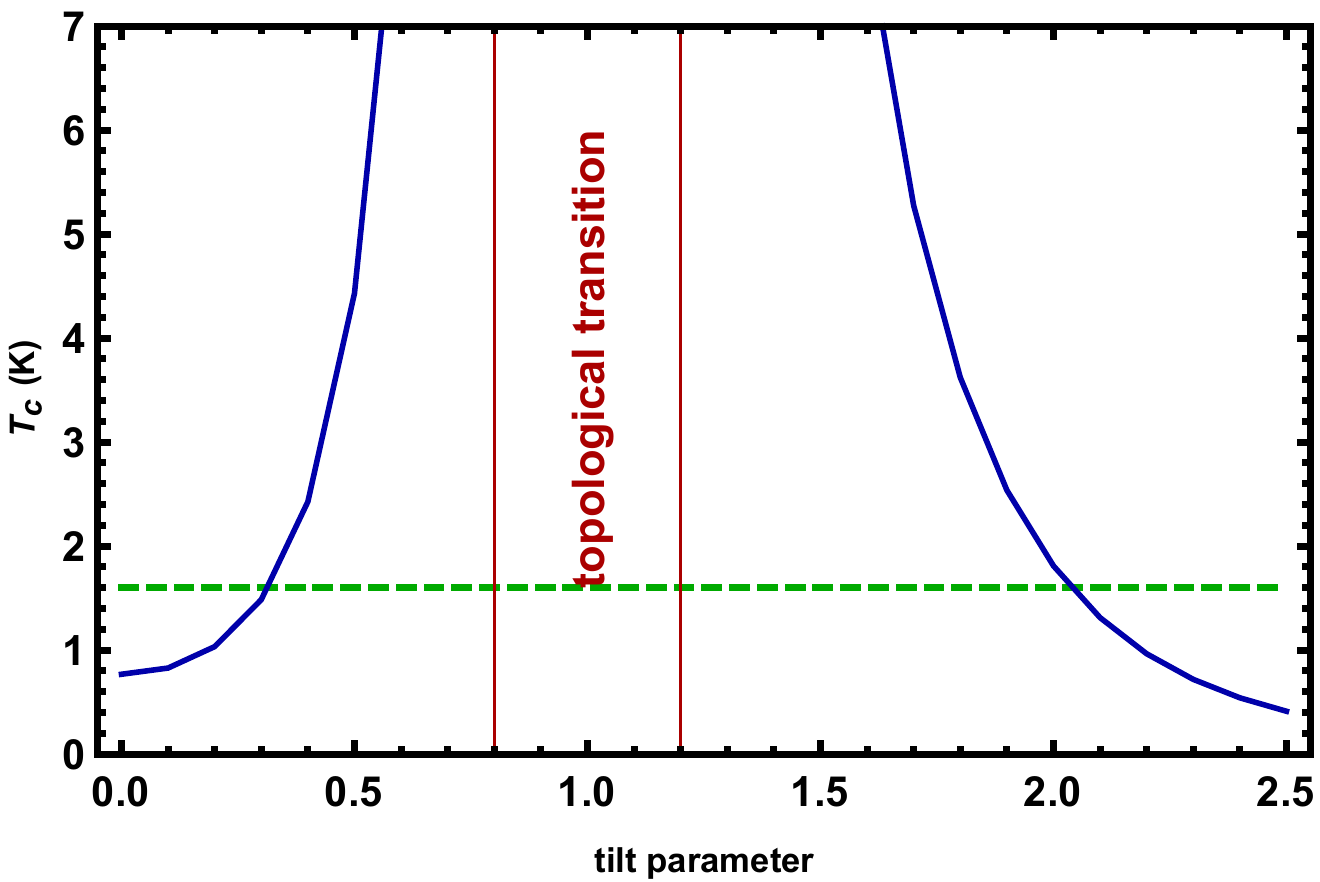} \vspace{-0.5cm}
\caption{Critical temperature as a function on the Dirac cone tilt parameter
$\protect\kappa $ plotted for electron-electron strength $\protect\lambda %
_{0}=0.2$, Effective mass in the third direction $z$ is $m_{z}=m_{e},$the
chemical potential $\protect\mu =25\Omega $ while $\Omega =100K$ is a cut of
phonon frequency.  Green dash line marks the critical temperature of the $%
PdTe_{2}$ sample from Ref. \protect\cite{Leng}.}
\end{figure}

\subsubsection{The gradient and the cubic terms of the GL equation}

Non-diagonal components $\ $of the second derivative tensor $C_{ik}$ are
zero due to the reflection symmetry in $p_{y}$ direction , when the cone
tilt vector $\mathbf{w}$ is directed along the $x$ axis. Using Eqs.(\ref{Eq.
39}) and (\ref{GF}), we obtain for diagonal components.

\begin{equation}
C_{zz}=\frac{3\hbar ^{2}\sqrt{T_{c}}}{4\pi m_{z}\mu ^{3/2}f}%
\sum\limits_{\omega }\int \frac{\varepsilon d\varepsilon \sqrt{\varepsilon
_{z}}d\varepsilon _{z}d\varphi }{Z}\left\{ 4\varepsilon ^{2}\left( \overline{%
\omega }^{2}+\Phi ^{2}\right) +\left( \overline{\omega }^{2}+\Phi
^{2}\right) ^{2}+2\Phi ^{2}\varepsilon ^{2}+\varepsilon ^{4}\right\} ,
\label{CZZ}
\end{equation}%
where $Z=\left( \overline{\omega }^{2}+\left( \Phi -\varepsilon \right)
^{2}\right) ^{2}\left( \overline{\omega }^{2}+\left( \Phi +\varepsilon
\right) ^{2}\right) ^{2}$. In the Dirac directions,\
\begin{eqnarray}
C_{xx} &=&\frac{3v^{2}\hbar ^{2}}{8\pi T^{2}\overline{\mu }^{3/2}f}%
\sum\limits_{\omega }\int \varepsilon d\varepsilon d\varphi \frac{%
d\varepsilon _{z}}{Z\sqrt{\varepsilon _{z}}}  \label{cXX} \\
&&\cdot \left(
\begin{array}{c}
+\left( 2\Phi \varepsilon \kappa \cos \varphi +\varepsilon ^{2}\cos 2\varphi
+\Phi ^{2}-\overline{\omega }^{2}\right) ^{2}+ \\
+\left( 2\kappa \overline{\omega }\varepsilon \sin \varphi \right)
^{2}+8\left( \varepsilon \sin \varphi \right) ^{2}\left( \kappa \Phi
+\varepsilon \cos \varphi \right) ^{2}+ \\
+\overline{\omega }^{2}\left( \overline{\mu }-\varepsilon _{z}\right) ^{2}+8%
\overline{\omega }^{2}\left( \kappa \Phi +\varepsilon \cos \varphi \right)
^{2}+ \\
+2\left( \kappa \Phi ^{2}+2\varepsilon \Phi \cos \varphi +\kappa \varepsilon
^{2}\right) ^{2}+2\overline{\omega }^{4}\kappa ^{2}- \\
-4\overline{\omega }^{2}\left( \kappa ^{2}\Phi ^{2}+2\kappa \Phi \varepsilon
\cos \varphi +\kappa ^{2}\varepsilon ^{2}\right) \
\end{array}%
\right)   \notag
\end{eqnarray}

\bigskip

\begin{eqnarray}
C_{yy} &=&\frac{3v^{2}\hbar ^{2}}{4\pi T^{2}\overline{\mu }^{3/2}f}%
\sum\limits_{\omega }\int \varepsilon d\varepsilon d\varphi \frac{%
d\varepsilon _{z}}{\sqrt{\varepsilon _{z}}}Z^{-1}  \label{CYY} \\
&&\cdot \left[
\begin{array}{c}
\left( \overline{\omega }^{2}-2\varepsilon ^{2}\sin ^{2}\varphi -\Phi
^{2}+\varepsilon ^{2}\right) ^{2}+4\left( \left( \varepsilon ^{2}\sin
\varphi \cos \varphi \right) ^{2}+\overline{\omega }^{2}\Phi ^{2}\right) \\
+4\left( \varepsilon \sin \varphi \right) ^{2}\left( \overline{\omega }%
^{2}+\Phi ^{2}\right)%
\end{array}%
\right] \text{.}
\end{eqnarray}

The coherence lengths are defined by the relations\
\begin{equation}
\xi _{x}^{2}=C_{xx},\xi _{y}^{2}=C_{yy}\text{, }\xi _{z}^{2}=C_{zz}\text{.}
\label{ksi1}
\end{equation}%
After after summation over Matsubara frequencies the integration over
momenta were performed numerically in a wide range of tilt parameter $\kappa
$ and presented in Fig. 3.

\begin{figure}[t]
\centering
\includegraphics[width=16cm]{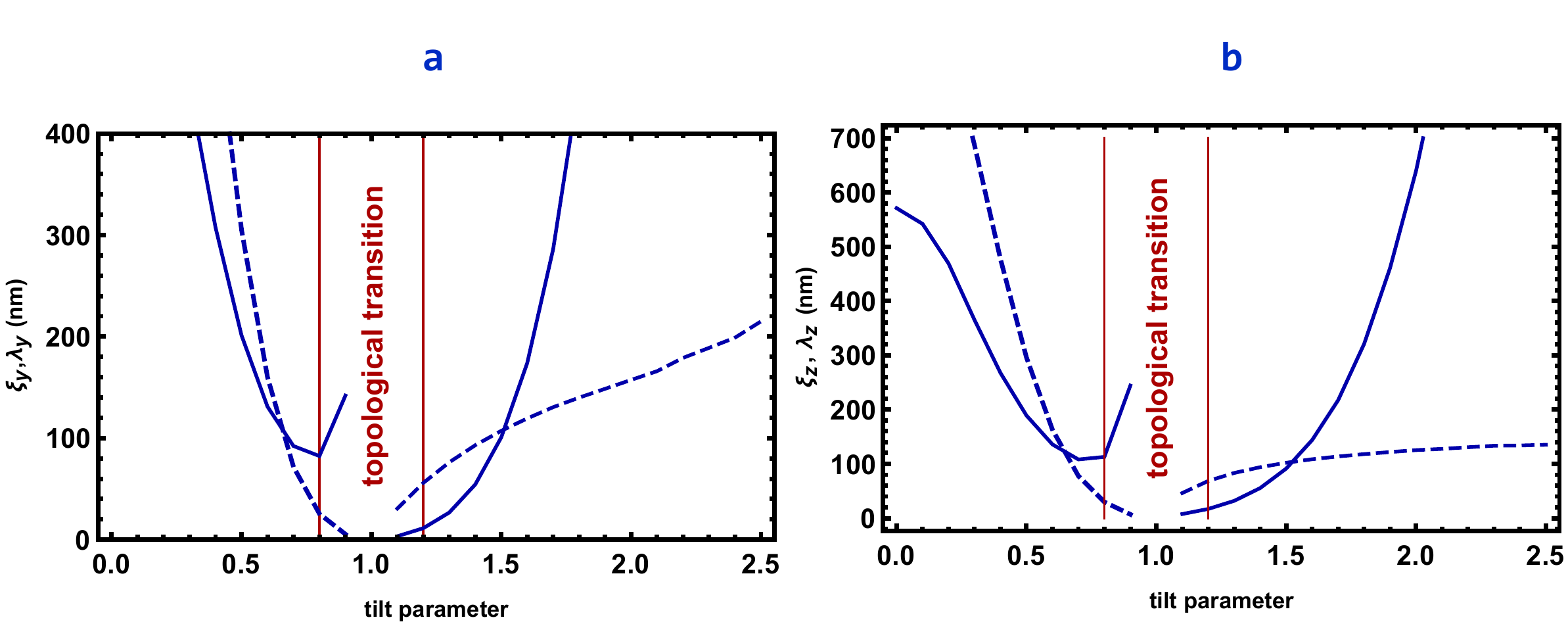} \vspace{-0.5cm}
\caption{Coherence\ length $\protect\xi $ (bold lines) and penetration
lengths (dashed lines) $\protect\lambda \protect\sqrt{2}$ as functions on
the Dirac cone tilt parameter $\protect\kappa $. Same parameters as in
Fig.2. }
\end{figure}

Substituting GF from Eq.(\ref{GF}) into Eq.(\ref{Eq.40}), one can express
the cubic term in the form

\begin{equation}
\beta =\ \frac{3\ \ }{4\pi \overline{\mu }^{3/2}T^{2}f}\sum\limits_{\omega
}\int \frac{d\varepsilon _{z}}{\sqrt{\varepsilon _{z}}}d\varphi \varepsilon
d\varepsilon \frac{\left[ \varepsilon ^{2}+\overline{\omega }^{2}+\Phi ^{2}%
\right] \left[ \varepsilon ^{2}+\overline{\omega }^{2}+\Phi ^{\prime 2}%
\right] }{\left[ \overline{\omega }^{2}+\left( \Phi -\varepsilon \right) ^{2}%
\right] \left[ \overline{\omega }^{2}+\left( \Phi +\varepsilon \right) ^{2}%
\right] \left[ \overline{\omega }^{2}+\left( \Phi ^{\prime }-\varepsilon
\right) ^{2}\right] \left[ \overline{\omega }^{2}+\left( \Phi +\varepsilon
\right) ^{2}\right] }\text{,}  \label{Betta}
\end{equation}%
where $\Phi =$\bigskip $\overline{\mu }-\varepsilon _{z}-\kappa \varepsilon
\cos \varphi ,\Phi ^{\prime }=\overline{\mu }-\varepsilon _{z}+\kappa
\varepsilon \cos \varphi $.

\subsection{Free GL energy for 3D DSM superconductor and penetration depths
in London limit.}

Effects of the external magnetic field are accounted for by the minimal
substitution, $\mathbf{\nabla }\rightarrow \mathbf{D}=\mathbf{\nabla }-\frac{%
2ei}{c}\mathbf{A}$ in the GL equation Eq.(\ref{GL17}) due to gauge
invariance.$\ $The GL equation in the presence of magnetic field allows the
description of the magnetic response to homogeneous external field. We start
from the strong field that destroys superconductivity.

Density of superconducting currents is obtained by the variation of the\
free energy functional including the magnetic energy,
\begin{equation}
F=\int d^{3}r\left\{ D_{0}\left( \mu \right) f\left( \xi _{ii}^{2}\left\vert
D_{i}\Delta \right\vert ^{2}-\tau \left\vert \Delta \right\vert ^{2}+\frac{%
\beta }{2}\left\vert \Delta \right\vert ^{4}\right) +\frac{\left( \nabla
\times \mathbf{A}\right) ^{2}}{8\pi }\right\} \text{,}  \label{energy3D}
\end{equation}%
with respect to components of the vector potential:\
\begin{equation}
\mathbf{J}_{i}=D\left( \mu \right) \frac{2ei}{\hbar }\xi _{ii}^{2}\Delta
\left( \mathbf{r}\right) D_{i}\Delta ^{\ast }\left( \mathbf{r}\right) +c.c.%
\text{.}  \label{j}
\end{equation}%
Within the London approximation, with magnetic field parallel to the $x$
direction (see Fig.1), taking the order parameter in the form $\Delta \left(
\mathbf{r}\right) =\Delta e^{i\varphi \text{ }}$, one obtains ,

\begin{equation}
\mathbf{J}_{i}=\frac{4e}{\hbar }D\left( \mu \right) \xi _{ii}^{2}\Delta
^{2}\left( \partial _{i}\varphi -\frac{2e}{c\hbar }A_{i}\right) \text{.}
\label{j1}
\end{equation}

The London penetration lengths in our case of DSM with parabolic dispersion
relation along $z$ axis are:
\begin{equation}
\lambda _{z}^{2}\left( T\right) =\frac{c^{2}\hbar ^{2}}{32\pi e^{2}D\left(
\mu \right) \xi _{y}^{2}\Delta ^{2}}\text{, }\lambda _{y}\left( T\right)
=\lambda _{z}\left( T\right) \frac{\xi _{y}}{\xi _{z}}\text{,}
\label{lambdas}
\end{equation}%
where $\Delta ^{2}=\tau /\beta $. Substituting DOS from Eq.(\ref{DOS}) one
obtains,
\begin{equation}
\lambda _{z}^{2}\left( T\right) =\frac{3\pi ^{2}\hbar ^{3}v^{2}c^{2}\hbar
^{2}\beta }{8\pi e^{2}\xi _{y}^{2}\sqrt{2m_{z}}\mu ^{3/2}f\tau },
\label{lamdaz}
\end{equation}%
and presented in Fig.3.

The Abrikosov parameter is isotropic despite large anisotropies:

\begin{equation}
\kappa _{z}^{A}=\kappa _{y}^{A}=\frac{\lambda _{z}}{\xi _{z}}=\frac{c\hbar }{%
e\xi _{y}\xi _{z}\Delta }\sqrt{\frac{1}{32\pi D\left( \mu \right) }}\ \text{.%
}  \label{Eq.69}
\end{equation}

\subsection{ Critical magnetic fields.}

Thermodynamic critical field for kind I superconductors is given by

\begin{equation}
H_{c}^{2}\left( 0\right) =8\pi F_{s}=4\pi D_{0}\left( \mu \right) f\Delta
^{2}=\frac{4\sqrt{2m_{z}}\mu ^{3/2}f^{2}}{3\pi \hbar ^{3}v^{2}\beta }\text{.}
\label{Hc}
\end{equation}%
The upper critical magnetic field $H_{c2}$ in kind I superconductors is
defined the overcooled critical field \cite{De Gennes}. It can be calculated
as usual from the linear part of the GL equation,

\begin{equation}
\left( -\xi _{ii}^{2}D_{i}-\tau \right) \Delta =0\text{,}  \label{GL1}
\end{equation}%
as the lowest eigenvalue of the linear operator (including the magnetic
field). Representing the homogeneous magnetic field in the $x$ axis
direction in the Landau gauge, $A=H\left( 0,0,y\right) $, \bigskip one
obtain near $T_{c}$ as%
\begin{equation}
H_{c2}\left( T\right) =H_{c2}\left( 0\right) \tau ,\   \label{Hc2}
\end{equation}%
where the zero temperature intercept magnetic field is $H_{c2}\left(
0\right) =\Phi _{0}/2\pi \xi _{z}\xi _{y}$.

\section{Discussion and Conclusions}

Magnetic properties of Dirac (Weyl) semi - metals superconductors with
"hybrid" dispersion relation of the electrons (Dirac in $x-y$ plane and
parabolic in $z$ direction) at low temperatures were derived from a
microscopic phonon mediated two - band pairing model via the Gorkov
approach for the (singlet) order parameter. Microscopically derived Ginzburg
- Landau effective theory was used to determine microscopically anisotropic
coherence length, the penetration depth, Fig.3, determining the Abrikosov
parameter for a such materials. It is found that generally strongly second
kind superconductivity in Dirac semimetals becomes first kind especially in
type II WSM. It was shown that relatively large Fermi energy is crucial for
existence of the kind one superconductivity effectively reducing the
Abrikosov parameter $\kappa ^{A}$ separating superconductors in two groups
with different magnetic properties. In DSM superconductors of first kind
both the thermodynamic field $H_{c}\left( T\right) $ and upper critical
field $H_{c2}\left( T\right) $ that takes a role of the supercooling field
is calculated.

Main results of the paper are presented in Figs.2-3 where solid and dashed
curves related to the coherence superconducting length and magnetic
penetration depths $\lambda \sqrt{2}$ correspondingly. Figures 3 demonstrate
that in the Type I phase of the DSM superconductivity is of the kind two $%
\left( \kappa ^{A}>1/\sqrt{2}\right) $ while in the phase Type II the DSM it
overcomes to the kind II superconducting state. Our results applied to the
DSM superconductor $PdTe_{2}$ and related systems. In particular, the
superconductor $PdTe_{2}$ was recently classified as a Type II Dirac
semimetal with magnetic measurements confirmed that $PdTe_{2}$ was a first
kind superconductor with $T_{c}=1.64$ $K$ and the thermodynamic critical
field of $H_{c}(0)$ $=13.6$ $mT$ \ (intermediate state under magnetic field
is typical to a first kind superconductor, as demonstrated by the
differential paramagnetic effect \cite{Leng}). Experimentally measured
effective Abrikosov parameter $\kappa ^{A}=0.13$ takes place at the
magnitude of cone tilt parameter $\kappa =2$ where $T_{c}=1.64K$ (see Fig. 3
and Eq.(\ref{Hc})).The temperature dependence of the thermodynamic magnetic
field is in agreement with results of our theory.

\bigskip

\bigskip \textit{Acknowledgements.}

We are grateful to N.L. Wang for valuable discussions. B.R. was supported by
NSC of R.O.C. Grants No. 103-2112-M-009-014-MY3 and is grateful to School of
Physics of Peking University and Bar Ilan Center for Superconductivity for
hospitality. The work of D.L. also is supported by National Natural Science
Foundation of China (No. 11274018 and No. 11674007).\newpage

\appendix

\section{Gorkov equations in integral form}

\bigskip Gorkov equations Eq.(\ref{FGE}) can be presented in an integral
form:

\begin{equation}
g_{\epsilon \kappa }\left( \mathbf{r,r}^{\prime }\ \omega \right) =\mathrm{g}%
_{\epsilon \kappa }^{1}\left( \mathbf{r}-\mathbf{r}^{\prime },\omega \right)
-\int \mathrm{g}_{\epsilon \theta }^{1}\left( \mathbf{r}-\mathbf{r}^{\prime
\prime },\omega \right) \Delta _{\theta \phi }^{\ast }\left( \mathbf{r}%
^{\prime \prime }\right) f_{\phi \kappa }^{+}\left( \mathbf{r}^{\prime
\prime }\mathbf{,r}^{\prime },\omega \right) ;  \label{A1}
\end{equation}%
\begin{eqnarray}
f_{\beta \kappa }^{+}\left( \mathbf{r,r}^{\prime },\omega \right)  &=&\int
\mathrm{g}_{\beta \alpha }^{2}\left( \mathbf{r-r}^{\prime \prime \prime
},-\omega \right) \Delta _{\alpha \epsilon }^{\ast }\left( \mathbf{r}%
^{\prime \prime \prime }\right) \cdot   \label{A2} \\
&&\cdot \left[ \mathrm{g}_{\epsilon \kappa }^{1}\left( \mathbf{r}^{\prime
\prime \prime }-\mathbf{r}^{\prime },\omega \right) -\int \mathrm{g}%
_{\epsilon \theta }^{1}\left( \mathbf{r}^{\prime \prime }-\mathbf{r}^{\prime
\prime \prime },\omega \right) \Delta _{\theta \phi }^{\ast }\left( \mathbf{r%
}^{\prime \prime }\right) f_{\phi \kappa }^{+}\left( \mathbf{r}^{\prime
\prime }\mathbf{,r}^{\prime },\omega \right) \right]   \notag
\end{eqnarray}

\bigskip\ Expanding in small order parameter $\Delta ,$ one obtains:

\begin{eqnarray}
\Delta \left( \mathbf{r}\right)  &=&\frac{g^{2}T}{2}\sum\limits_{\omega
}\int \mathbf{dr}^{\prime \prime \prime }\left[
\begin{array}{c}
\left[ \mathrm{g}_{21}^{2}\left( \mathbf{r-r}^{\prime \prime \prime }\right)
\mathrm{g}_{21}^{1}\left( \mathbf{r}^{\prime \prime \prime }-\mathbf{r}%
\right) \right] \sigma _{12}^{x}\sigma _{12}^{x} \\
+\left[ \mathrm{g}_{11}^{2}\left( \mathbf{r-r}^{\prime \prime \prime
}\right) \mathrm{g}_{22}^{1}\left( \mathbf{r}^{\prime \prime \prime }-%
\mathbf{r}\right) \right] \sigma _{21}^{x}\sigma _{12}^{x}+ \\
\left[ \mathrm{g}_{12}^{2}\left( \mathbf{r-r}^{\prime \prime \prime }\right)
\mathrm{g}_{12}^{1}\left( \mathbf{r}^{\prime \prime \prime }-\mathbf{r}%
\right) \right] \sigma _{21}^{x}\sigma _{21}^{x} \\
+\left[ \mathrm{g}_{22}^{2}\left( \mathbf{r-r}^{\prime \prime \prime
}\right) \mathrm{g}_{11}^{1}\left( \mathbf{r}^{\prime \prime \prime }-%
\mathbf{r}\right) \right] \sigma _{12}^{x}\sigma _{21}^{x}%
\end{array}%
\right] \Delta \left( \mathbf{r}^{\prime \prime \prime }\right)   \label{A3}
\\
&&-\int \mathbf{dr}^{\prime \prime \prime }\mathbf{dr}^{\prime \prime }%
\mathbf{dr}_{3}\Pi \left( \mathbf{r,r}^{\prime \prime \prime }\mathbf{,r}%
^{\prime \prime }\mathbf{,r}_{3}\right) \Delta _{\theta \phi }^{\ast }\left(
\mathbf{r}^{\prime \prime }\right) \Delta _{\alpha \epsilon }^{\ast }\left(
\mathbf{r}^{\prime \prime \prime }\right) \Delta _{\zeta \epsilon }^{\ast
}\left( \mathbf{r}_{3}\right)   \notag
\end{eqnarray}

where $\Pi \left( \mathbf{r,r}^{\prime \prime \prime }\mathbf{,r}^{\prime
\prime }\mathbf{,r}_{3}\right) =\mathrm{g}_{\beta \alpha }^{2}\left( \mathbf{%
r-r}^{\prime \prime \prime }\right) \mathrm{g}_{\epsilon \theta }^{1}\left(
\mathbf{r}^{\prime \prime }-\mathbf{r}^{\prime \prime \prime }\right)
\mathrm{g}_{\phi \zeta }^{2}\left( \mathbf{r}^{\prime \prime }\mathbf{-r}%
_{3}\right) \mathrm{g}_{\epsilon \kappa }^{1}\left( \mathbf{r}_{3}-\mathbf{r}%
\right) $

\section{ Calculation of the normal GF}

Normal Green function obeyed the equations (\ref{Eq.12},\ref{Eq.14}). First
four GF are calculated from the equation

\bigskip
\begin{equation}
L_{\gamma \beta }^{1}\mathrm{g}_{\beta \kappa }^{1}\left( \mathbf{r-r}%
^{\prime }\right) =\delta ^{\gamma \kappa }\delta \left( \mathbf{r-r}%
^{\prime }\right) ,  \label{B1}
\end{equation}%
where $L_{\gamma \beta }^{1}=\left[ \left( i\omega +\mu +i\mathbf{w\nabla }%
_{r}\right) \delta _{\gamma \beta }+\left( -iv\sigma _{\gamma \beta }^{i}\
\nabla _{r}^{i}\right) \right] $ by performing Fourier transform for
different pseudo-spin indexes. In particular for $\gamma =1,\kappa =1\ $it
reads in momentum representation\bigskip \
\begin{eqnarray}
\ \left( i\omega +\mu -\mathbf{wp}\right) g_{11}^{1}\left( \mathbf{p}\right)
+v\left( \ \mathbf{p}^{x}-i\ \mathbf{p}^{y}\right) g_{21}^{1}\left( \mathbf{p%
}\right) &=&1;  \label{B2} \\
\left( i\omega +\mu -\mathbf{wp}\right) g_{11}^{1}\left( \mathbf{p}\right)
+vp\left( \ \cos \varphi -i\sin \varphi \right) g_{21}^{1}\left( \mathbf{p}%
\right) &=&1.  \notag
\end{eqnarray}%
The rest of the normal GF may be obtained by the same method. The second
group of the normal Green functions obey the equations $L_{\gamma \beta
}^{2}g_{0\beta \kappa }^{2}\left( \mathbf{r-r}^{\prime }\right) =\delta
^{\gamma \kappa }\delta \left( \mathbf{r-r}^{\prime }\right) \ $with $%
L_{\gamma \beta }^{2}$ defined in Eq.(\ref{Eq.12}) are obtained by the same
method. \bigskip The GF obtained after solution of these equations are:

\begin{eqnarray}
g_{22}^{1}\left( \mathbf{p}\right) &=&z^{\ast -1}\left( i\omega +\mu -%
\mathbf{wp}\right) ;\text{ \ }g_{12}^{1}\left( \mathbf{p}\right) =-z^{\ast
-1}vpe^{-i\varphi }  \label{B3} \\
g_{11}^{1}\left( \mathbf{p}\right) &=&z^{\ast -1}\left( i\omega +\mu -\
\mathbf{wp}\right) ;\text{ }g_{21}^{1}\left( \mathbf{p}\right) =-z^{\ast
-1}vpe^{i\varphi }  \notag \\
g_{11}^{2}\left( \mathbf{p}\right) \ &=&z^{-1}\ \left( -i\omega +\mu -%
\mathbf{wp}\right) ;\text{ }g_{12}^{2}\left( \mathbf{p}\right)
=-z^{-1}vpe^{i\varphi }  \notag \\
g_{22}^{2}\left( \mathbf{p}\right) &=&z^{-1}\left( -i\omega +\mu -\ \mathbf{%
wp}\right) ;\text{ \ }g_{21}^{2}\left( \mathbf{p}\right)
=-z^{-1}vpe^{-i\varphi }\ ;  \notag \\
z &=&\left( -i\omega +\mu -\mathbf{wp}\right) ^{2}-\left( vp\right) ^{2},
\notag
\end{eqnarray}%
where $\mathbf{p}$ is the 2D momentum and $\varphi $ is the azimuthal angle
in the $p_{x},p_{y}$ plane.\newpage

\

\end{document}